\documentclass[final,5p,times,twocolumn]{elsarticle}

\makeatletter

\everymath{\displaystyle} 

\usepackage{mathrsfs}

\usepackage{amssymb}
\usepackage{amsmath}
\usepackage{mathtools}
\usepackage{subfig}
\usepackage{array}
\usepackage{soul}
\usepackage{xcolor}
\usepackage{algorithm,algpseudocode}
\usepackage{hyperref}

\usepackage{wrapfig}

\usepackage[utf8]{inputenc} 
\usepackage[T1]{fontenc}

\usepackage[colorinlistoftodos, textsize=scriptsize]{todonotes} 
\usepackage[normalem]{ulem}

\begin{document}

\begin{frontmatter}

\title{Lifetime Optimization of Dense Wireless Sensor Networks Using Continuous Ring-sector Model }

\author[label1]{Arouna Ndam Njoya \corref{cor1}}
\ead{ndanjoa@gmail.com}
\cortext[cor1]{Corresponding author}
\author[label2]{Christopher Thron}
\author[label3]{Marah Nana Awa}
\author[label3,label4]{Ado Adamou Abba Ari }
\author[label3]{Abdelhak Mourad Gueroui}

\address[label1]{Department of Computer Engineering, University Institute of Technology, University of Ngaound\'er\'e, P.O. Box 455 Ngaound\'er\'e, Cameroon}
\address[label2]{ Department of Mathematics, Texas A \& M University-Central Texas, Killeen, USA}
\address[label3]{LaRI Lab, University of Maroua, P.O. Box 814 Maroua, Cameroon}
\address[label4]{DAVID Lab, Universit\'e Paris-Saclay, University of Versailles Saint-Quentin-en-Yvelines, 45 Avenue \'Etats-Unis 78035 Versailles cedex, France}

\begin{abstract}

Wireless sensor networks (WSNs) are becoming increasingly utilized in applications that require remote collection of data on  environmental conditions. In particular dense WSNs are emerging as an important  sensing platforms for the Internet of Things (IoT). WSNs are able to generate huge volumes of raw data, which require network structuring and efficient collaboration between nodes to ensure efficient transmission.  In order to reduce the amount of data carried in the network, data aggregation is used in WSNs to define a policy of data fusion and compression. In this paper, we investigate a model for data aggregation in a dense {WSN} with a single sink. The model divides a circular coverage  region centered at the sink into patches which are intersections of  sectors of concentric rings, and data in each patch is aggregated at a single node before transmission. Nodes only communicate with other nodes in the same sector. Based on these assumptions, we formulate a linear programming problem to maximize system lifetime by minimizing the maximum proportionate energy consumption over all nodes. Under a wide variety of conditions, the optimal solution employs two transmissions mechanisms: direct transmission, in which nodes send information directly to the sink; and  stepwise transmission, in which nodes transmit information to  adjacent nodes. 
An exact formula is given for the proportionate energy consumption rate of the network.  Asymptotic forms of this exact solution are also derived, and are verified to agree with the linear programming solution.
We investigate three strategies for improving system lifetime: nonuniform energy and information density; iterated compression; and modifications of rings. We conclude that  iterated compression has the biggest effect in increasing system lifetime. 
\end{abstract}


\end{frontmatter}

\section{Introduction}
\label{intro}
The term "sensor" may be applied to any device that serves as a bridge between the physical world and the digital world \cite{karim2016sensor}. Wireless sensor networks (WSN) are spatial distribution of sensors that communicate over radio links for the detection of events and the transmission of the collected data to the sink in a given region \cite{njoya2020hybrid}. The sensors in WSN's are typically characterized by limited computing power, as well as limited power capacity \cite{ari2016power,njoya2015evolutionary}. Applications of WSN's include environmental monitoring, health, smart agriculture, smart cities, and so on \cite{sahitya2016wireless,njoya2017efficient,gondchawar2016iot,ali2019q}.

The applications of sensor networks have experienced significant growth with the advent of the Internet of things (IoT). IoT includes devices with embedded sensors and a network interface that allows it to interconnect physical and virtual worlds \cite{zhu2015green,karim2016sensor}. IoT applications lead to massive connectivity that produces large amounts of data. The collection and  the processing of these data require the implementation of robust methods that maintain the availability of network resources \cite{boubiche2018big,deebak2020hybrid}. 

Data aggregation is a mechanism used in WSNs to reduce the amount of data that is being transmitted over the network, which has the benefits of reducing power consumption (thus extending network lifetime), reducing transmission delays, and increasing security of  data transmission \cite{boubiche2018big,ari2020three}. 

Many studies use a network configuration composed of a sink deployed at the center of network, and source nodes distributed throughout the area of interest. The source nodes are responsible for sensing and transmitting the information collected to the  sink using single-hop or multi-hop transmission \cite{chen2008adaptive, elhoseny2017dynamic}. An alternative configuration consists in integrating multiple sinks in the network \cite{xue2005maximizing, fitzgerald2018energy}. The architecture where the mobility is included into the network has also been the subject of studies in \cite{konstantopoulos2009effective, gupta2014energy}. Since source nodes generally spend more energy during communication, another form of network model is to divide it into small hexagonal cells \cite{li2014latency}. In the same vein, network models integrating fog computing and the cloud are also used to limit long distance communications between nodes \cite{bonomi2012fog}. The definition of the network model also requires specifying the homogeneous or heterogeneous character of nodes \cite{qing2006design}.

Whatever the configuration adopted, typically the source nodes in the network are generally partitioned into clusters \cite{ari2016power,ari2020three,sahoo2020particle}. The idea is to deploy source nodes around the cluster heads (CHs), which merge data collected from the cluster members and transmit it to the sink. Note here that the transmission can be done directly from the CH to the sink or by multi-hop through other CHs \cite{al2019energy}.

The main challenge in partitioning the network into clusters lies in the specification of a scheme for election of CHs that optimizes the WSNs performance in terms of reducing latency and/or maximizing system lifetime . Election schemes that have been used previously include random selection, as well as algorithms  that take into account the residual energy of the nodes, the election threshold probability, the distance between the CH and its members, the distance between CH and the sink, or some combination of these mechanisms \cite{heinzelman2000energy, al2019energy}.

Furthermore, since data aggregation aims to improve the use of network resources, many studies have focused on network lifetime extension \cite{gbadouissa2020hgc,njoya2016optimization,heinzelman2000energy,kalpakis2003efficient,han2019novel, al2019energy,labraoui2013reactive}. Other topics that have been investigated include  utilization of data compression in order to reduce the overload of nodes as well as those of transmission links \cite{xiang2011compressed, yao2014edal,wu2016data, gilbert2018trust, xu2019big}, redundant data elimination \cite{ccam2006energy,harb2014k}, latency and packet transmission delay minimization \cite{lu2004adaptive,huang2012evolution, liu2013approximation,li2014latency, dong2015joint}, routing path selection \cite{hussain2006hierarchical, iqbal2013advanced, asemani2015learning, randhawa2017data,myoupo2018fault}, interference reduction between nodes \cite{liu2013approximation, li2014latency}, and data forwarding with confidentiality and integrity preservation \cite{ozdemir2011integrity, lin2012cdama, sicari2012dydap, lopez2017evolving}. Solution methods include, linear programming \cite{liu2013approximation, yagouni2015contribution}, approximation algorithms \cite{liu2013approximation}, machine learning \cite{ghate2018machine, ullah2020intelligent, ullah2020efficient, lu2018self, donta2020congestion} and techniques from cryptography as elliptic curves \cite{ozdemir2011integrity, lin2012cdama}.



In this paper, a data aggregation approach for network lifetime extension and data fusion is investigated for a WSN where sensors are densely (but not necessarily uniformly) distributed around a single sink in a circular region. Unlike many previous studies that seek to identify individual CHs and/or  transmission paths based on the locations of individual source nodes, we assume a dense, continuous model in which the sources node locations are not specifically identified. 
In order to understand the factors that influence network lifetime, we propose a simple network structure that uses rings and sectors to divide the region into patches, such that sensor data in each patch is aggregated at a single node (which may change over time), before transmission towards the sink either directly or via other nodes. 
Lifetime optimization is formulated as a  linear programming problem, and solutions are computed for a variety of network parameters to determine characteristics of the optimal information flow. Based on these characteristics, a theoretical expression for optimal flow is derived under very general conditions on the system parameters, as well as theoretical and approximate expressions for the system depletion rate (which is inversely proportional to the system lifetime). These approximate expressions give insight into possible strategies for improving system lifetime, which are also verified against the linear programming solutions. 
 
The remainder of this paper is organized as follows. Section~\ref{sec2} discusses  related work on network lifetime optimization and data aggregation.  Section~\ref{sec3} presents our ring-sector model, together with linear constraints on the model. Section~\ref{sec4} formulates the WSN lifetime optimization problem, and gives theoretical analysis including exact and approximate solutions for a wide class of system parameters. Section~\ref{sec5} describes the numerical simulations which confirm and clarify the theoretical results. Section~\ref{sec6} summarizes the research and draws final conclusions.

\section{Related work}
\label{sec2}

The main focus of this paper is on schemes for aggregation and transmission of the information that is gathered at source nodes and sent to the sink, such as to  minimize power consumption and extend the lifetime of the WSN. Before presenting out results, we first  review previous research concerned with extending  lifetime of WSNs. 

The low-energy adaptive clustering hierarchy (LEACH) protocol for data aggregation proposed by \cite{heinzelman2000energy, heinzelman2000application} is widely used to reduce energy dissipation of WSNs. The main idea behind this protocol is to avoid direct transmission of the information to the sink, since this leads to high energy consumption. For this purpose, LEACH organizes sensor nodes around cluster heads (CH). The CHs perform data collection, and compression of the information from all other sensors in the cluster, and then transmits the aggregated information to the sink \cite{kalpakis2003efficient, qing2006design}. The LEACH protocol is performed in two phases: the cluster setup phase,  in which CHs are selected; and the steady state phase,  where aggregation and further communication with the sink is accomplished. These two phases are iterated, so that the CHs vary from iteration to iteration.

LEACH does have some drawbacks. Since LEACH selects CHs randomly, it does not take into account the residual energy  of cluster members. Another problem with LEACH is that the number of CH nodes is not fixed as a result of stochastic selection. To improve CH selection,  a modification of LEACH was proposed in \cite{al2019energy} which uses three-level CH selection based on a genetic algorithm. This modified approach takes residual energy of cluster members into account, as well as the distance from the CH to the sink.  

A hierarchical cluster-based routing (HCR) data aggregation method  where the network configuration of CHs is sent to source nodes from the sink is designed in \cite{hussain2006hierarchical}. The method includes a genetic algorithm that  generates cluster distributions according to a given number of transmissions which optimizes the energy of the network. Even if the proposed evolutionary algorithm improves the energy consumption of the network, the stability of the network is affected due to the abstract design of the fitness function, as pointed out in \cite{bara2012new}.


A hexagonal structure based data aggregation clustering algorithm is proposed in \cite{mammu2015cross} as a means to prolong network lifetime. The first election of CH is performed randomly. To be re-elected, the CH broadcasts an election message to sensor nodes within one hop distance of its position, in which it specifies its residual energy, location and the average distance to nodes in his vicinity (in one hop). All other cluster members also send election requests in the same way. The final CH is one of the nodes that is close to the CH and has high residual energy and minimum average distance to his neighbors. Compare to LEACH protocol, this cross-layer cluster-based energy-efﬁcient protocol (CCBE) proposed gives better network lifetime extension. However this method requires significant overhead \cite{han2019novel}.

A learning automata-based aggregation (LAG) which combine the residual energy of sensor, the number of hops from source node to sink and the dynamic context of a network was proposed in \cite{asemani2015learning} to intelligently find the best path to forward data to the sink. To cope with network context change, each node is equipped with learning automata which help him to select the right hop for data transferring to the sink. This means that each sensor node adapts the context for sending information according to changes in the network \cite{randhawa2017data}. By introducing intelligence the proposed method aims to enhance data aggregation and network load balancing simultaneously. However the analysis shows that this mixture increase the delay time of packet forwarding.

Different macine learning tools of varying sophistication have been used to optimize data aggregation  in WSNs. An overview of such methods is given in \cite{ghate2018machine, ullah2020intelligent}.  For instance, a three-layer  neural  network  mixed with  Mahalanobis distance was propose in \cite{ullah2020efficient} for extreme learning in order to increase the number of live nodes in the network and clustering accuracy. The intelligence provided by machine learning allows to include context awareness into data aggregation using some powerful tools as  Q-learning \cite{lu2018self, donta2020congestion}. The main challenge in the use of machine learning methods lies in the computation time involved.

For heterogeneous WSNs, the distributed energy-efficient clustering (DEEC) algorithm  was proposed in \cite{qing2006design}. To maintain a higher number of live nodes  over time,  each sensor node determines  itself at each election round if it can be CH or  not using his residual energy, the  average energy of the  network and the weighted probability threshold. Although this protocol reduces the network energy expenditure, the election strategy does not include multi-hop transmission between clusters and dynamic context of a node during the selection phase of CHs as indicated in  \cite{elhoseny2017dynamic}. Various alternative  implementations of DEEC can be found in \cite{singh2017energy}.  

An extension of LEACH protocol called advanced LEACH (Ad-LEACH) which may be applied in heterogeneous routing context is proposed in \cite{iqbal2013advanced}. The idea is to group the nodes into static clusters of small sizes in order to limit the energy consumed during broadcast transmission over a long distance. Another reason is to reduce the complexity of cluster management. Two types of nodes compose the network: advanced nodes and the normal nodes. Advanced nodes has more energy than the normal nodes. Despite the improvement brings out by Ad-LEACH, there still a challenge in generating network structure that uniformly controls the energy expenditure of each sensor node, as highlighted in \cite{elhoseny2014balancing}. Network structure control is developed in \cite{elhoseny2014balancing} using a genetic algorithm which implements two operations for chromosome validation to check if each node within the chromosome is available and has sufficient energy before applying the fitness function. These operations are needed to avoid introducing infeasible solutions into the population.

The above studies all deal with the case of a single sink. 
To handle  the case of multiple sinks, \cite{xue2005maximizing} utilizes  linear programming and multicommodity flow.  Alternatively, reference \cite{rajagopalan2006data} uses forest data aggregation to handle the case of multi-sink: however, performance analysis shows that the energy gain tends to decrease as the number of sinks increase .
Reference \cite{fitzgerald2018energy} uses  mixed-integer programming and  min-max optimization, and is implemented on a network architecture that relies on fog computing, such that several nodes are introduced in order to route data from the source node to the desired destination. A survey of data aggregation algorithms for WSNs with multiple sinks may be found in \cite{curry2016survey}.


In order to reduce the number of comparisons of similar data transmitted to the sink, the $k$-means  prefix frequency filtering (KPFF) data aggregation was proposed in \cite{harb2014k}. The method is a combination of the $k$-means algorithm and the  prefix frequency filtering (PFF) algorithm previously proposed in \cite{bahi2012optimized}. The first step of the approach is to use the $k$-means algorithm at the level of data, and then apply the PFF to perform clustering. In this approach, $k$-means can be seen as preprocessing which helps the algorithm to reduce the computation time required to eliminate redundant data. The proposed scheme decreases the percentage of data that is forwarded to the sink, consequently reducing the data latency as well as the energy consumption of the network. However, as pointed out in \cite{harb2017comparison}, PFF which is  the main part of this method provides poor energy conservation.

Redundant data is reduced in \cite{ccam2006energy} using pattern codes for better bandwidth utilization and energy efficiency. In fact, to accommodate node failures, sensor nodes are commonly deployed at high density. This generally leads to overlapping sensing radii which causes the sending of similar data to the CHs by multiple sensor nodes. Instead of checking redundant data at the CH level, the proposed energy-efficient secure pattern based data aggregation (ESPDA) generates pattern codes corresponding  to the characteristics of data sensed which allows the coordination of sleeping and active states of overlapping sensing ranges.  As a result, a reduced amount of encrypted data is transmitted to the cluster head, leading to better bandwidth utilization and network lifespan saving. However the global gain offering by the compression code and the related intra-cluster communication cost were not be evaluated in the proposed study, as they were in\cite{zheng2010distributed, wu2016data}. Moreover, this aggregation algorithm does not address the problem of errors in data transmission and the data compression ratio \cite{roy2012secure, zhang2018multi}.

Data compression is one of the features used in data aggregation to reduce traffic on links between nodes. Along these lines, \cite{xiang2011compressed} first gives a mixed-integer programming formulation of compressed sensing aggregation,  then proves its NP-completeness, and finally proposes a greedy heuristic for its resolution. Four data collection mechanisms are utilized: non-aggregated data collection, where no compression is performed to initial raw data;  plain compressed sensing aggregation, which requires each link to carry $k$ samples; lossy data aggregation, which extracts statistically certain quantities of information in data transmitted  by source nodes; and hybrid compressed sensing aggregation, which combines the non-aggregated and plain compressed sensing mechanisms. However, the proposed scheme induces a high computational load, and compressed sensing reconstruction errors are not investigated in this study. The latter deficiency is addressed in the compression scheme of \cite{yao2014edal}. A robust compression model that includes data prediction as in \cite{wu2016data, gilbert2018trust}. A summary of the literature on data compression in WSNs may be found in \cite{xu2019big}.

\section{System model}
\label{sec3}
In this paper, we consider a sensor network 
with a single sink which functions as sink node, and with sensors densely distributed in a circular region surrounding the sink.
The model  takes into account the possibility that sensor densities and residual energy may vary throughout the region.

In the following subsections we describe in detail the geometry of the model, and the mathematical constraints governing the model's behavior.

\subsection{Geometry of the ring model}

\begin{figure}[!h]
\begin{center}
\includegraphics[width=8cm]{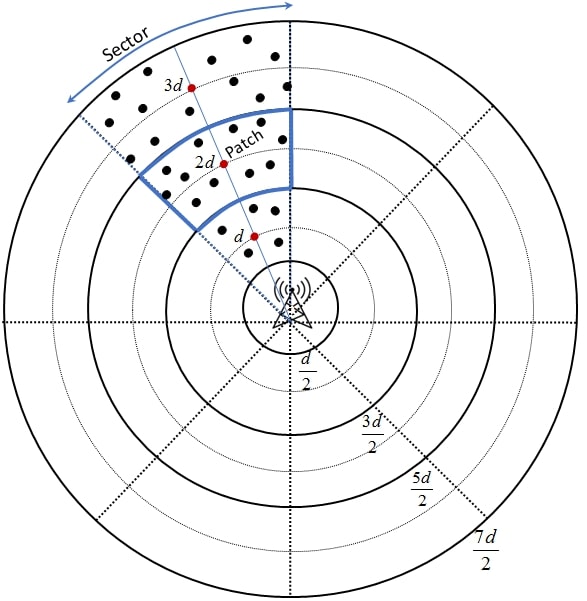}
\end{center}
\caption{ Simplified sectorized ring model for the dense sensor network.}
\label{ring_model}
\end{figure}

The ring model for the dense sensor network is shown in Figure~\ref{ring_model}. The sink is located at a stationary position in the center of the network and sensor nodes are randomly distributed within a circular region centered at the sink.
The region's area is divided into  $N$ concentric annuli (rings) centered on the sink node. Each ring is divided into $S$ sectors, thus dividing the entire circular region into $N\times S$ patches shaped like curved rectangles (see Figure~\ref{ring_model}), plus a circle  at the center. We suppose that all of the sensors within each patch send their information to a single CH node at the center of the patch, which then transmits the information towards the sink. This node may itself be a sensor: if so, then nodes may change over time in order to equalize the power load among sensors in the patch.     
We also suppose that no information is transmitted between sectors: instead, information is passed between the central nodes in each sector. 
Sector boundaries may also rotate over time, so that specific patches change over time. However, the geometry is preserved, and at any given time there is no information transfer between sectors.
Since the different sectors don't interact, we can solve for the power and information flow for each sector independently.  

\subsection{Line-of-nodes representation of ring sectors}\label{sec_LON}

We shall suppose that the inner radius of the $j$'th ring is $r_j = (j - 1/2)d$, $j=1,\ldots N$. so that patch centers are located at radii $d,2d,3d, \ldots$.  In this case, the model for each sector can be  represented schematically as a sequence of $N+1$ equally-spaced nodes with interval $d$. These nodes transmit information between themselves, such that eventually all the information is transmitted to node $0$. 

We define additional notation as follows (note that index $j$ ranges from 1 to $N$, and the index $i$ ranges from 0 to $N$): 
\begin{itemize}
\item 
$a_j \ge 0$ is the information per time associated with node $j$ which is to be transmitted the sink;
\item  
$b_j$ is the compression applied to all information sent by node $j$ ($0 < b_j \le 1$);
\item
$c_j \ge 0$ is the current energy capacity left in node $j$;
\item
$t_{ij} > 0$ is the power cost to node $j$ of transmitting a unit of information from node $j$ to node $i$ (which may also include processing power cost). In the following, we will assume that $t_{ij}$ is a convex function of $|i-j|$;
\item 
$x_{ij} \ge 0$ is the amount of information per unit time that flows from node $j$ to node $i$.
\end{itemize}
The system is represented schematically in Figure~\ref{ring_model}.

\subsection{Constraint equations for line-of-nodes model}
Given the model representation described in Section\ref{sec_LON}, the equations governing the behavior of the information transfer in the line-of nodes as follows. We suppose that  the values $a_j, b_j$, and $t_{ij}$ remain approximately constant over the time interval of interest.  The total information  processed by  node $j$ per time is equal to the information originating from $j$ (i.e. $a_j$) plus the information per time that flows into node $j$ from other nodes. This information is  compressed at rate $b_j$, and then subsequently transmitted to other nodes. This relationship between inflowing and outflowing information for each node gives an equality constraint for each node $j$, $j=1,\ldots N$ as follows:
\begin{equation}
\label{eq_consringinfo}
\sum_{i=0}^{N} x_{ij} = b_j \left(a_j+\sum_{i=1}^{N} x_{ji}\right), j=1,\ldots,N
\end{equation}

 We also put a nonnegativity constraint on the information flows $x_{ij}$, and prohibit flows from nodes to themselves:
\begin{equation}
\label{eq_consnonnulldata}
x_{ij}\geq 0;~~ x_{ii}=0, \qquad 0 \le i,j \le N.
\end{equation}

In the case where $b_j=1$ for all $j$, we may add together the $N$ equations \eqref{eq_consringinfo} and obtain:
\begin{equation}
\begin{aligned}\label{eq_infoBalance}
\sum_{j=1}^N \sum_{i=0}^N x_{ij} &= \sum_{j=1}^N a_j+ \sum_{j=1}^N \sum_{i=1}^{N} x_{ji}\\
\implies
\sum_{j=1}^N  x_{0j} &= \sum_{j=1}^N a_j.
\qquad (\text{when } b_j = 1~ \forall j)
\end{aligned}
\end{equation}

Equation \eqref{eq_infoBalance} guarantees that all information eventually flows to the sink node at 0, when no compression is used.   However, when compression is used (i.e. $\min_{1 \le j \le N} b_j < 1$), then it is possible for information to compress down to 0 by shuttling back and forth between two nodes. To prevent this, we require that all information should move towards the sink node, i.e. 
\begin{equation}\label{eq_compress}
\min_{1\le j \le N} b_j < 1 \implies x_{ij} = 0, 1 \le j <i \le N. 
\end{equation}
This requirement is reasonable for typical systems, because in practice information is sent towards the sink node and not away from it.

\section{System optimization}\label{sec4}

Given the constraints expressed in \eqref{eq_consringinfo}, \eqref{eq_consnonnulldata}, \eqref{eq_compress}, we may now formulate different optimization problems for the system.  
In this section will consider two separate optimization problems: (1) minimize total power consumption; (2) maximize network lifetime.

\subsection{Minimization of total power consumption}
First, we consider the problem of minimizing the total power consumption of the network, which may be formulated as follows:
\begin{equation}\label{totPowerMin}
\textrm{Minimize } P_{tot} := \sum_{i=0}^N\sum_{j=1}^N t_{ij}x_{ij} \textrm{ Subject to constraints \eqref{eq_consringinfo},\eqref{eq_consnonnulldata},\eqref{eq_compress}}.
\end{equation}
 
This problem may be solved by considering information from each node $j$ independently, and finding the transmission path from $j$ to $0$ that minimizes total power. Given the transmission path $n_M, n_{M-1}, \ldots n_0$ where $n_0 = 0$ then the power required to send $1$ unit of information per time  along that path is:
\begin{equation}\label{eq_powerCostPerUnit}
\textrm{Total per-unit power for node } n_M = t_1 \sum_{m=1}^{M} \left( {\textstyle \prod_{\ell = m}^M b_{n_\ell}}\right),
\end{equation}
where $t_1$ is the energy required to transmit one unit of information between adjacent nodes. 
. In the usual case, the transmission coefficient $t_{ij}$ is a convex function of $|i-j|$.  In this case, the minimum-energy path from $j$ to $0$ will be $j,j-1,\ldots 0$, so that the energy required for node $j$ to send information at rate $a_j$ is
\begin{equation}\label{eq_powerCostInfo_j}
\textrm{Total power for info. from node } j = t_1 a_j \sum_{m=1}^{j} \left( {\textstyle \prod_{\ell = m}^{j} b_{\ell}}\right) .
\end{equation}
It follows that the minimum total power expenditure for all nodes is:

\begin{equation}\label{eq_totPowerCost}
P_{tot,min} = t_1 \sum_{j=1}^N  \left( a_j  \sum_{m=1}^{j} \left( {\textstyle \prod_{\ell = m}^{j} b_{\ell}}\right) \right) 
\end{equation}
This solution leads to very uneven power consumption among nodes, since all information from  nodes greater than $j$ passes through node $j$. The  power expenditure for node $j$ may be computed explicitly as follows:
\begin{equation}\label{eq_powerCostNode_j}
\textrm{Power expended by node } j = t_1  \sum_{m=j}^{N} a_m \left( {\textstyle \prod_{\ell = j}^{m} b_{\ell}}\right) .
\end{equation}

In the case where information compression is constant for all nodes  (i.e. $ b_{j}:=\beta ~\forall j$, expression \eqref{eq_powerCostNode_j} simplifies to
\begin{equation}\label{eq_simplePowerCostNode_j}
\begin{aligned}
&\textrm{Power expended by node } j \text{ at constant compression} \\
&\qquad = t_1   \sum_{m=0}^{N-j} a_{j+m} \beta^{m+1} .
\end{aligned}
\end{equation}

In the case where sensors are uniformly distributed, then $a_j$ is proportional to the area of a patch in the $j$'th ring shown in Figure~\ref{ring_model}, which is proportional to $j$. This implies $a_j = ja_1$, which leads to 
\begin{equation}\label{eq_simplePowerCostNode_j_2}
\begin{aligned}
&\textrm{Power expended by node } j \text{ at constant compression and density}\\
& \qquad= 
\begin{cases}
\frac{\beta t_1 a_1}{(1-\beta)^2}\left[\beta + j(1-\beta) - \beta^{N-j+1}((1-\beta)N+1)\right] & (\beta<1) \\
\frac{t_1 a_1}{2}\left( N(N+1) - j(j-1) \right)  & (\beta = 1)
\end{cases}\\
\end{aligned}
\end{equation}
Thus when no compression is present, the power consumption of nodes in the first ring is roughly proportional to the square of the number of rings.

\subsection{Maximization of network lifetime}\label{sec_maxLife}
In the case where nodes' energy capacities are limited, the network can no longer function when any node's energy reserve is depleted. Maximizing the remaining network lifetime is equivalent to minimizing the maximum depletion rate, where the depletion rate of a node is given by the node's power consumption divided by the node's remaining energy capacity.
Using the terminology defined above, the power consumption of node $j$ is given by  $\sum_{i=0}^N t_{ij}x_{ij}$, so the depletion rate of node $j$ is  $\sum_{i=0}^N t_{ij}x_{ij}/c_j$.  The lifetime of node $j$ at this level of power consumption is  $\left(\sum_{i=0}^N t_{ij}x_{ij}/c_j\right)^{-1}$.  Thus  system lifetime may be maximized by solving the following optimization problem:
\begin{equation}\label{eq_propPower}
\begin{aligned}
\text{Minimize } &\Phi, \text{ subject to constraints \eqref{eq_consringinfo}, \eqref{eq_consnonnulldata},\eqref{eq_compress} and}\\
& \Phi \geq \sum_{i=0}^{N} \frac{t_{ij}}{c_j}x_{ij}, j=1,\ldots,N
\end{aligned}
\end{equation}
We will refer to $\Phi$ in \eqref{eq_propPower} as the system's depletion rate, since the system lifetime is $\Phi^{-1}$.
 This solution to \eqref{eq_propPower} may be readily found using linear programming. In order to explore the dependence of solutions on system parameters, we make the following assumptions on node parameters' dependence on location:
\begin{equation}\label{eq_abc}
\begin{aligned}
a_j &= k_a j^\alpha;\\
b_j & = \beta,~0 <\beta \le 1 ;\\
c_j &= k_c  j^\gamma,
\end{aligned}
\end{equation}

where $\alpha, \beta,\gamma, k_a, k_c$ are constants.
 We also assume a power law for transmission power with exponent $\lambda$:
\begin{equation}\label{eq_lambda}
\textstyle
t_{ij} = k_t \cdot dist(i,j)^{\lambda},
\end{equation}
where $dist(i,j)$ is the distance from node $i$ to node $j$.  Assuming that the nodes are uniformly spaced with spacing $d$, we have:
\begin{equation}\label{eq_dist}
dist(i,j) = d \cdot |i-j|.
\end{equation}
In the simulations, distances are normalized so that  $d=1$.
In addition, the proportionality constants $k_a, k_c,$ and $k_t$ are chosen so as to present results in a scale-invariant manner. Specifically, values are set as follows:
\begin{equation}\label{eq_ka_kc}
k_a = \frac{\pi N^2 (d+1/2)^2}{\sum_{j=1}^N j^\alpha}; \quad k_c = \frac{\pi N^2 (d+1/2)^2}{ \sum_{j=1}^N j^\gamma};\quad k_t = 1.
\end{equation}  
The normalizations $k_a, k_c$ for $a_j$ and $c_j$ respectively guarantee that the  information density (information per area) and power capacity density (power capacity per area) over all nodes are both equal to 1.  Due to the linearity of the model, solutions for other values of $k_a, k_c$, and $k_t$ retain the same functional form, and can be obtained by simple scaling.

As will be shown in Section~\ref{sec3}, for a wide range of the system parameters $\alpha, \beta, \gamma, \lambda$ defined in \eqref{eq_abc} and \eqref{eq_lambda},  the  solutions to the optimization problem \eqref{eq_propPower} utilize only direct and stepwise transmission: in other words, $x_{ij} = 0$ unless $i=0$ or $i = j-1$.  Furthermore, in these solutions it was found that all nodes have the same depletion rate in these solutions. In the following, the stepwise flow from node $j$ to node $j-1$ is denoted as $y_j$,  
and the depletion rate is denoted as  $\Phi$.

\begin{figure}[!h]
\begin{center}
\includegraphics[width=6.5cm]{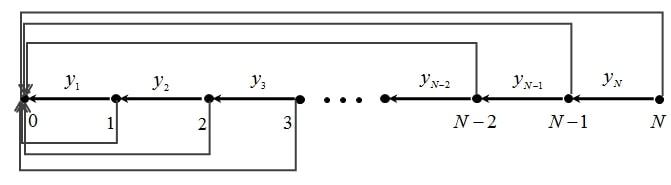}
\end{center}
\caption{Nodes information flow}
\label{infoFlowNodes}
\end{figure}

\begin{figure}[!h]
\begin{center}
\includegraphics[width=6.5cm]{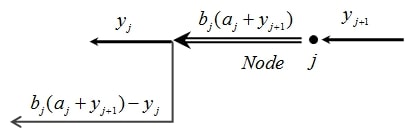}
\end{center}
\caption{Single node information ﬂow processing}
\label{singleNodeInfo}
\end{figure}

In Fig. \ref{singleNodeInfo}, we illustrate how the information flows in a single node. The figure may be summarized as follows:

\begin{itemize}
	\item The information into to node $j$ is $y_{j+1}$
	\item The information out from node $j$ is $b_j(a_j+y_{j+1})$. As we say previously $b_j$ is the compression factor.
	\item The information passed from node $j$ to node $j-1$ is equal to $y_j$. 
	\item The information passed from node $j$ directly to node 0 (the sink) is $b_j(a_j+y_{j+1})-y_j$.
\end{itemize}

The two outflowing branches in Figure~\ref{singleNodeInfo} are associated with power flows $y_j$ and  
$(b_j(a_j+y_{j+1})-y_j)j^{\lambda}$ respectively, so that the total power consumption for node $j$ is

\begin{equation}\label{eq_nodePower}
y_jd^\lambda+(b_j(a_j+y_{j+1})-y_j)(jd )^{\lambda}~~(j=1\ldots N-1).
\end{equation}

Since we are assuming that the depletion rate $\Phi$ is equalized for all nodes, it follows that $c_{j}\Phi$ is the power consumption at node $j$.  Setting \eqref{eq_nodePower} equal to $c_{j}\Phi$ and rearranging gives 

\begin{equation}
\label{eq_nodePower2}
y_j(1-j^{\lambda}) + y_{j+1}b_j j^{\lambda}= c_j d^{-\lambda} \Phi - b_j a_j j^{\lambda}~~(j=1\ldots N).
\end{equation}
The first $N-1$ of these equations may be expressed in matrix form as:
\begin{equation}\label{eq_mx1}
\begin{aligned}
&\begin{bmatrix}
b_{1} & \cdots & \cdots &\cdots & -c_1 d^{-\lambda} \\
(1-2^{\lambda}) & 2^{\lambda}b_{2} & 0 &0 &-c_2 d^{-\lambda} \\
0 & \ddots & \ddots & 0 &\vdots  \\
\vdots  & \ddots  &  \ddots& \ddots& \vdots  \\
0 & \cdots &  0 &(1-N^{\lambda})& -c_N d^{-\lambda} 
\end{bmatrix}
\quad
\begin{bmatrix}
 y_{2} \\
y_{3} \\
 \vdots  \\
y_{N} 
\end{bmatrix}\\
&\qquad \quad
=
\begin{bmatrix}
 c_1 d^{-\lambda}\Phi -b_{1}a_{1} \\
c_2 d^{-\lambda}\Phi -b_{2}a_{2}2^{\lambda}  \\
 \vdots  \\
 \vdots  \\ 
c_N d^{-\lambda}\Phi -b_{N}a_{N}N^{\lambda}
\end{bmatrix}
\end{aligned}
\end{equation}
With the following definitions:
\begin{equation}\label{eq_pqwDef}
\begin{aligned}
p_{j}&=\frac{c_{j}}{b_{j}}(jd)^{-\lambda}\\
q_{j}&=\frac{(1-j^{-\lambda})}{ b_{j}}\\
w_{j} &=p_{j}\Phi-a_{j}.
\end{aligned}
\end{equation}
we may rewrite \eqref{eq_mx1} as:

\begin{equation}
\begin{bmatrix}
1 & 0 & \cdots &\cdots & 0 \\
-q_{2} & 1 &\ddots  & &\vdots \\
0 & \ddots & \ddots & \ddots &\vdots \\
\vdots  & \ddots  & \ddots& \ddots &0  \\
0 & \cdots  & 0 &-q_{N-1}& 1 
\end{bmatrix}
\quad
\begin{bmatrix}
 y_{2} \\
y_{3} \\
 \vdots  \\
 \vdots  \\
y_{N} 
\end{bmatrix}
=
\begin{bmatrix}
 w_{1}\\
w_{2} \\
 \vdots  \\
 \vdots  \\ 
w_{N-1}
\end{bmatrix}
\end{equation}
This subdiagonal matrix equation may be solved by forward substitution:
\begin{equation}
y_j = w_{j-1} + q_{j-1}y_{j-1},
\end{equation}
which gives

\begin{equation}
\begin{aligned}
y_2 &= w_1;\\
y_3 &= w_{2} + q_{2}y_{2} = w_{2} + q_{2}w_1;\\
y_4 &= w_{3} + q_{3}w_{2}  + q_3q_2w_1;\\
y_5 &= w_{4} + q_{4}w_{3}  + q_4q_3w_2 + q_4q_3q_2w_1;\\
&\vdots
\end{aligned}
\end{equation}
In general we may write:

\begin{equation}\label{eq_yj}
y_j = w_{j-1} + \sum_{k=1}^{j-2}\left( {\textstyle \prod_{\ell=k+1}^{j-1}q_\ell} \right) w_k
\end{equation}
Since $y_{N+1}=0$, we have from \eqref{eq_nodePower2} that

\begin{equation} \label{eq_yN}
\begin{aligned}
&y_N(1-N^{\lambda}) =c_N \Phi+b_N a_N N^{\lambda}\\
\implies &y_N = -w_N/q_N
\end{aligned}
\end{equation}
Setting\eqref{eq_yj} with $j=N$  equal to  \eqref{eq_yN} gives:

\begin{equation}\label{eq_wN}
0 = w_N +   \sum_{k=1}^{N-1}\left( {\textstyle \prod_{\ell=k+1}^{N}q_\ell} \right) w_k
\end{equation}
Substituting the expressions for $w_j$ from   \eqref{eq_pqwDef} yields
\begin{equation}\label{eq_tPhi}
\Phi  = \frac{a_N +   \sum_{k=1}^{N-1}\left( {\textstyle \prod_{\ell=k+1}^{N}q_\ell} \right) a_k}{p_N +   \sum_{k=1}^{N-1}\left( {\textstyle \prod_{\ell=k+1}^{N}q_\ell} \right) p_k},  
\end{equation}
which may be written very succinctly as:
\begin{equation}\label{eq_PhiSimple}
\Phi   = \frac{\vec{a} \cdot \vec{\phi}}{\vec{p} \cdot \vec{\phi}},
\end{equation}
where
\begin{equation}
\begin{aligned}
\vec{a} &:=\left[ a_{1}, a_{2}, \ldots, a_{N} \right]\\
\vec{p} &:= \left[ p_{1},p_{2}, \ldots, p_{N} \right]\\
\vec{\phi} &:= \left[ \phi_1,\ldots,\phi_N\right]
\end{aligned}
\end{equation}
and
\begin{equation}
\begin{aligned}
\phi_1 &:=1;\\
\phi_j &:=
\prod_{k=2}^{j} q_j^{-1} = \prod_{k=2}^{j} \frac{b_k }{1 - k^{-\lambda}} ~~( 2 \le j \le N).
\end{aligned}
\end{equation}
Given these expressions, we may also write the stepwise flow $y_j$ as
\begin{equation}
y_j = \phi_j^{-1}(\Phi \vec{p} - \vec{a})\cdot [\phi_1,\ldots \phi_j,\underbrace{0,\ldots, 0}_{N-j~\text{entries}}]).
\end{equation}
When the parameters $a_j, b_j, c_j$ have the functional form given by \eqref{eq_abc} we may find an approximate analytical expression for $\phi_j$ as follows. If we define
\begin{equation}
    Q(j) :=\prod_{k=2}^{j} (1 - k^{-\lambda}),
    \end{equation}
It follows that
\begin{equation}
\begin{aligned}
    &Q(j+1) = (1 - N^{-\lambda})Q(j) \\
    &\implies \frac{dQ}{dj} \approx 
    - j^{-\lambda}Q(j) \\
    &\implies  Q(j) \approx 
    Q_0\exp\left( \frac{j^{1-\lambda}}{\lambda - 1} \right)\\
    & \implies  \frac{1}{Q(j)} \approx Q_0\left(1 - \frac{j^{1-\lambda}}{\lambda-1}
    + \frac{1}{2}\left(\frac{j^{1-\lambda}}{\lambda-1}
\right)^2
    \right).
\end{aligned}
    \end{equation}
From \eqref{eq_PhiSimple} we may obtain (using the definition of $p_j$ in \eqref{eq_pqwDef})
\begin{equation}\label{eq_PhiOrig}
\begin{aligned}
        \Phi &= \frac{\vec{a} \cdot \vec{\phi}}{\vec{p} \cdot \vec{\phi}} \\
        &\approx \beta d^\lambda \frac{k_a }{k_c} \frac{\sum_{j=1}^N  j^{\alpha}\beta^{j} \exp\left( \frac{j^{1-\lambda}}{\lambda - 1} \right) 
        }{\sum_{j=1}^N  j^{\gamma - \lambda}\beta^{j} 
        \exp\left( \frac{j^{1-\lambda}}{\lambda - 1} \right)
        }\\
        &\approx \beta d^\lambda\frac{\sum_{j=1}^N  j^{\gamma}}{\sum_{j=1}^N  j^{\alpha}} \frac{\sum_{j=1}^N  j^{\alpha}\beta^{j} \left(1 - \frac{j^{1-\lambda}}{\lambda-1}
    + \frac{1}{2}\left(\frac{j^{1-\lambda}}{\lambda-1}
\right)^2
    \right)}{\sum_{j=1}^N  j^{\gamma - \lambda}\beta^{j} \left(1 - \frac{j^{1-\lambda}}{\lambda-1}
    + \frac{1}{2}\left(\frac{j^{1-\lambda}}{\lambda-1}
\right)^2
    \right)},
\end{aligned}
\end{equation}
where we have replaced $k_a$ and $k_c$ with the expressions in \eqref{eq_ka_kc}.
It is clear from \eqref{eq_PhiOrig} that the relative sizes of $\alpha, \gamma$, and $\gamma-\lambda$ are key in determining the behavior of $\Phi$ as $N$ becomes large.  In the typical case, $\alpha=\gamma=1$ and $\lambda=2$, which implies that  the numerator of \eqref{eq_PhiOrig} grows much more rapidly than the denominator so that $\Phi$ increases rapidly with increasing $N$.

To further explore the behavior of $\Phi$ when $N$ is large, 
we may  
distinguish two practical situations:  low compression, corresponding to $\beta^N \approx 1$; and  high compression, corresponding to $N^{\nu}\beta^N \gtrapprox 0$. 
In the low-compression case where $\beta^N \approx 1$, we may  use $\beta^j \approx 1 - j(1-\beta)$ in \eqref{eq_PhiOrig} to convert all summed expressions to polynomials in $j$. We may also approximate sums with integrals: keeping up to first-order terms in $1-\beta^N$ and $N^{1-\lambda}$, we obtain:

\begin{equation}\label{eq_PhiApprox}
    \begin{aligned}
        &\Phi \approx \frac{\beta (\tilde{N}d)^{\lambda}}{\gamma+1} \cdot \left(1 
        - c_1 (1-\beta^N)
        - c_2\tilde{N}^{1 - \lambda}  \right) \cdot \\
        &\begin{cases}
            \frac{(\gamma - \lambda + 1)
            }
             {1 
             - c_3\tilde{N}^{1-\lambda} 
            -c_4 \tilde{N}^{-\gamma+\lambda-1}} 
             &(1  + \gamma- \lambda > 0); \\[1em]
                    \frac{1}{\log \tilde{N} 
            + c_5\tilde{N}^{1-\lambda}
           -c_6 \tilde{N}^{2-2\lambda} 
           - c_7}
            & (1  + \gamma- \lambda = 0).
        \end{cases}
    \end{aligned}
\end{equation}
where
\begin{equation}
    \begin{aligned}
    \tilde{N} &:= N + 0.5;\\
    c_1 &:=\frac{\alpha+1}{\alpha + 2}\\
    c_2 &:= \frac{(\alpha+1)}{(\lambda-1)(\alpha - \lambda+2)}\\
    c_3 &:= \frac{\gamma -\lambda +1}{(\lambda-1)(\gamma - 2\lambda + 2)}\\
    c_4 &:=  1 -  
             \frac{(\gamma -\lambda+1)2^{-2 + 2\lambda-\gamma}}{(\lambda-1)(\gamma - 2\lambda+2)}\\
    c_5 &:= \frac{1}{(\lambda-1)^2}\\
    c_6 &:=\frac{1}{4(\lambda-1)^3}\\
    c_7 &:= - \frac{2^{\lambda-1}}{(\lambda-1)^2}
           + \frac{4^{\lambda-2}}{(\lambda-1)^3} +\log 2.
    \end{aligned}
\end{equation}

In the high-compression case, the two sums involving $\beta^j$ in \eqref{eq_PhiOrig} both converge as $N \rightarrow \infty$. This means that $\Phi \sim N^{\gamma - \alpha}$ as $N \rightarrow \infty$, when $\beta < 1$.

 
\section{Simulation Results}\label{sec5}
\subsection{Overview}
In this section, we conduct several simulations in order to evaluate the proposed data aggregation schemes.  All simulations represent circular region where the sink is deployed in the center and the radius is known. 
First we vary the system exponents $\alpha, \gamma, \lambda$ and compression $\beta$ to characterize their effects on the optimal system solution's transmission characteristics. 
Next we investigate the effect of these parameters on the depletion rate (which determines system lifetime).    
Finally, we  investigate scaling properties as the number of rings $N$ and distance between rings $d$ are varied.

The system parameter ranges used in the simulations are shown in Table~\ref{tparam-sim}.

\begin{table}[ht!]
\begin{center}
 \centering
\caption{\label{tparam-sim}  Simulation parameters}
 \begin{tabular}{l|l|l|l}
 \hline
\textbf{Parameter}  & \textbf{Significance} & \textbf{Baseline} & \textbf{Range}\\
 &  & \textbf{value} &\\
\hline
 $\alpha$& Info. density exponent & 1 & $[0,3]$\\
$\beta$& Compression ratio & 1 & $[0.5,1]$\\
$\gamma$& Capacity exponent & 1& $[0,3]$\\
$\lambda$&Transmission exponent & 2& $[1.1,3]$\\
$N$& Number of rings & $20$&$5,...,20$\\
$d$& Ring spacing &1 &$[1/N,1]$\\
\hline
\end{tabular}
\end{center}
\end{table}

\subsection{Transmission characteristics' dependence on system parameters}

\begin{figure*}
  \centering
  \subfloat[System characteristics with  information capacity  exponent $\alpha=0,3$  and other parameters as in Table~\ref{tparam-sim} ]{\includegraphics[scale=0.5]{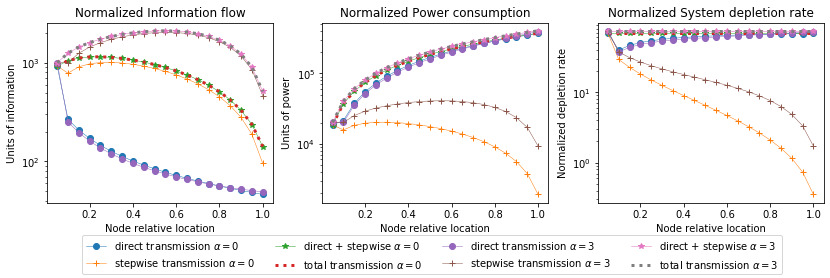}}\\
  \subfloat[System characteristics with  compression rate $\beta=0.5,1$ and other parameters as in Table~\ref{tparam-sim} ]{\includegraphics[scale=0.5]{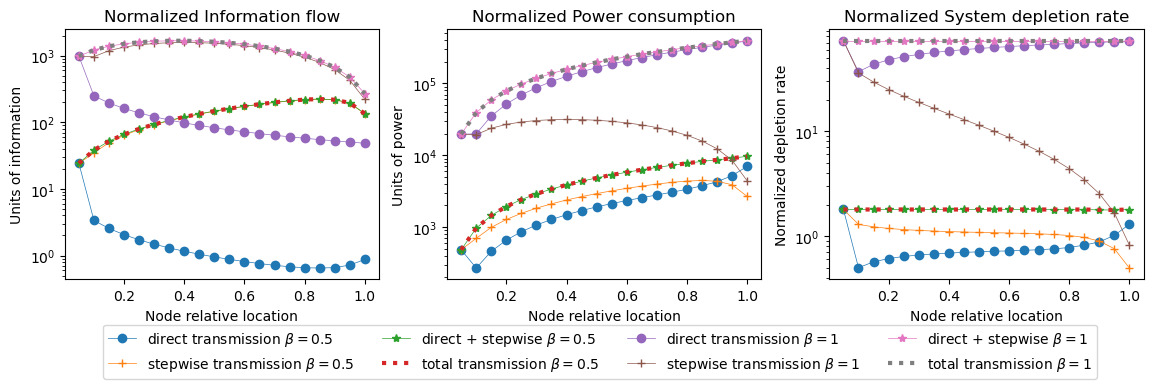}}\\
  \subfloat[System characteristics with  energy capacity  exponent $\gamma=0,3$  and other parameters as in Table~\ref{tparam-sim} ]{\includegraphics[scale=0.5]{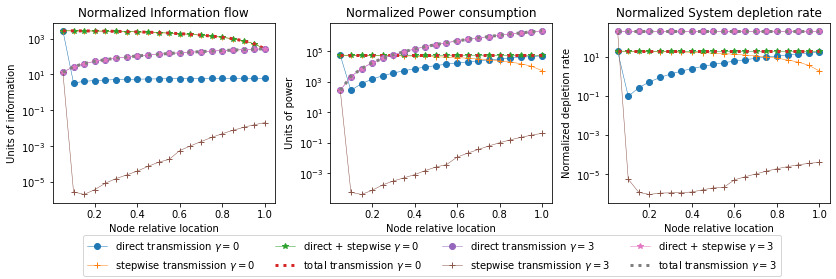}}\\
  \subfloat[System characteristics with  transmission power  exponent $\lambda=1.1,3$ and other parameters as in Table~\ref{tparam-sim} ]{\includegraphics[scale=0.5]{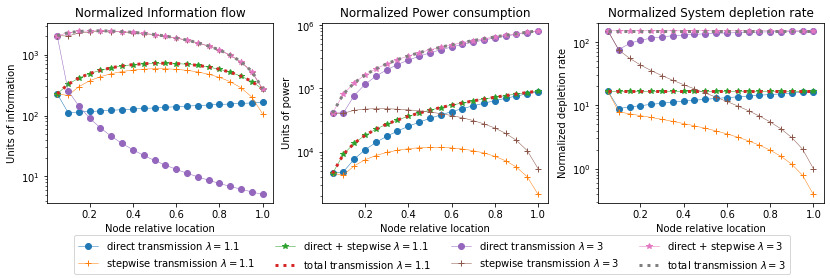}}
	  \caption{Normalized information flows, power consumption, and system depletion rates corresponding to direct, stepwise, and direct plus stepwise  transmission, for ranges of values of the parameters $\alpha,\beta, \gamma,$ and $\lambda$.} 
	  
  \label{direct-stepwise-total-power-flow}
\end{figure*}

In the first series of simulations, baseline values were first set for all parameters, and then individual parameters were varied one by one while keeping other parameters fixed. 
Baseline parameter values and parameter ranges are shown in Table \ref{tparam-sim}. Figure \ref{direct-stepwise-total-power-flow} shows the results of simulations. Figures (a), (b), (c), (d) correspond to four sets of simulations in which values of $\alpha, \beta, \gamma, \lambda$ respectively are varied from baseline values. Each set of simulations uses two different values of the varied parameter, as described in Table \ref{tparam-sim}. For each of the four sets of simulations, three graphs are given, The first graph (on the left) shows amount of direct and stepwise information transmitted by each of the $N$ nodes in the system. For node $j$, these values correspond to $x_{0j}$ and $x_{j-1,j}$  respectively as defined in Section \ref{sec_LON}. These values are plotted on the $y$ axis versus node relative position on the $x$ axis, where the relative position of node is given by $j/N$ (e.g. the farthest node is at relative position 1).
The second graph (center) shows direct and stepwise power transmitted by node $j$ for $1\le j \le N$, which corresponds to $t_{0j}x_{0j}$ and $t_{j-1,j}x_{j-1,j}$ respectively as defined in Section \ref{sec_LON}. The third graph (right) shows node depletion due to direct and stepwise information transmission for each node $j$ for $1\le j \le N$, which corresponds to $t_{0j}x_{0j}/c_j$ and $t_{j-1,j}x_{j-1,j}/c_j$ respectively as defined in Section \ref{sec_LON}. In each of the three graphs, the sums of direct plus stepwise values for each node  are also given.

It was observed  that in all optimal solutions simulated,
all transmission is either direct or stepwise:  in other words, $x_{ij} = 0$ unless $i=0$ or $i=j-1$.
This is reflected in the constant per-node depletion rate due to direct plus stepwise transmission observed in the third graph in all four series of graphs. This constant depletion rate is equal to the system depletion rate $\Phi$ defined in Section~\ref{sec_maxLife}. 

We may also notice consistent patterns in the information flow graphs (on the left) for all simulations. Directly transmitted information from node $j$ consistently decreases as the node's relative position increases. This is reasonable, since a larger relative position means that direct transmission has a higher power cost. In contrast, both stepwise information flow and total information flow consistently rise and then fall with increasing node index, where the maximum stepwise flow is typically reached between relative position 0.4--0.8. This is due to multiple aggregation along the line of nodes, so that node $j$ is not only transmitting its own information but also that of higher nodes. As the node index approaches 0, due to increasing direct transmission the information passed to lower nodes decreases, leading to a reduction in overall information transmission for small node indices.

In the following paragraphs, we will discuss the implications of Figures \ref{direct-stepwise-total-power-flow}(a),(b), (c), and (d) for the influence of parameters $\alpha, \beta, \gamma,$ and $\lambda$ respectively on system behavior. 

From Figures \ref{direct-stepwise-total-power-flow}(a) it is evident that the information density exponent $\alpha$ has little effect on the network's transmission characteristics for $0.5 \le \alpha \le 3$.  This is somewhat surprising, because a larger value of $\alpha$ is associated with larger information density that is far from the sink, and one would expect that more energy would be required to convey information to the sink, thus increasing depletion rate. However, this result validates \eqref{eq_PhiApprox}, which shows that $\Phi$ is nearly independent of $\alpha$ when $\beta = 1$. The main effect of increasing $\alpha$ within this range is to increase the stepwise transmission, while the increase in stepwise power consumption is offset by a decrease in power consumption from direct transmission. Note that although more information is sent stepwise than directly, the direct power consumption is much higher.  Recall that the power consumption is not equal for all nodes, because the power capacity increases linearly with node index since $\gamma = 1$.

Figures \ref{direct-stepwise-total-power-flow}(b) show the effect of data compression on system information and power flows. The information flow plot shows that with  $50\%$ compression, the direct information flow reduced by a factor of 10. Since most of the power consumption is due to direct information flow, this serves to reduce  power consumption (as well as depletion rate) by a factor of 10.  As in the first series of graphs, the stepwise information flow achieves its maximum at an intermediate node with relative location close to 0.5.

Figures \ref{direct-stepwise-total-power-flow}(c) show the effect of energy capacity exponent $\gamma$ on the system.
From this figure, we observe that $\gamma$ has great influence in stepwise transmission.  When $\gamma=3$, most of the energy capacity is contained in the nodes that are far from the sink, so they can expend much more energy and still preserve the same depletion rate as closer nodes. In this case, almost all information transmission is direct, as shown in the first graph in the row. However, when $\gamma=0$ there is much less energy capacity density far from the sink, and  the system relies mostly on stepwise transmission to compensate. 
Overall, the larger value of $\gamma$ gives rise to a depletion rate that is about 3 times larger, because direct transmission consumes more power than stepwise transmission.

Figures \ref{direct-stepwise-total-power-flow}(d) show the effect of the transmission power exponent $\lambda$ on the system. Larger values of $\lambda$ mean that direct transmission becomes increasingly costly compared to stepwise transmission, especially for nodes with high index.  This is reflected in figure that shows information flow, which indicates that direct transmission for $\lambda = 3$ is much larger than for $\lambda = 1.1$, especially for nodes with relative location near 1. When $\lambda = 1.1$, much of the information is sent directly, which also reduces the need for stepwise transmission, which has the effect of further lowering the power consumption. Altogether, the depletion rate for the system with $\lambda=3$ is about 10 times higher than the depletion rate for $\lambda = 1.1$.

\subsection{Effect of system parameters on system lifetime}

In this subsection we present several plots that display the effect of system parameters on depletion rate $\Phi$, which is inversely proportional to system lifetime. Several of the plots are heat maps that show how the magnitude of $\Phi$ varies as a function of two system parameters, thus elucidating interactions between parameters.

\begin{figure}[!h]
\begin{center}
\includegraphics[width=8cm]{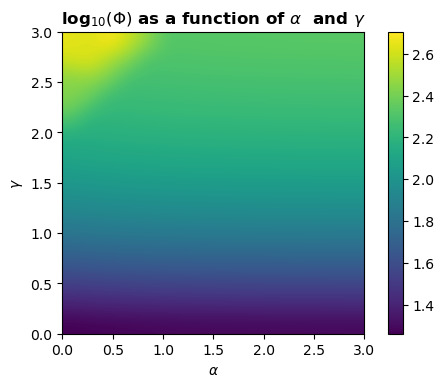}
  \caption{Log of system depletion rate $\log_{10}(\Phi)$ as a function of information density exponent $\alpha$ and energy capacity density exponent $\gamma$. }\label{depletion-rate-alpha-gamma}
 \end{center}
 \end{figure}

Figure~\ref{depletion-rate-alpha-gamma} shows how the information density exponent $\alpha$ and energy capacity density exponent $\gamma$ influence the depletion rate $\Phi$. When $\gamma < 2$, then $\alpha$ has very little effect on $\Phi$. On the other hand, $\Phi$ decreases steadily with decreasing $\gamma$ for all values of $\alpha$: for example, increasing $\gamma$ from 1 to 0 reduced $\Phi$ by a factor of about 3, independent of the value of $\alpha$. This shows that a smaller energy capacity exponent (corresponding to a higher energy capacity density near the sink) is associated with a longer system lifetime.

\begin{figure}[!h]
\begin{center}
\includegraphics[width=8cm]{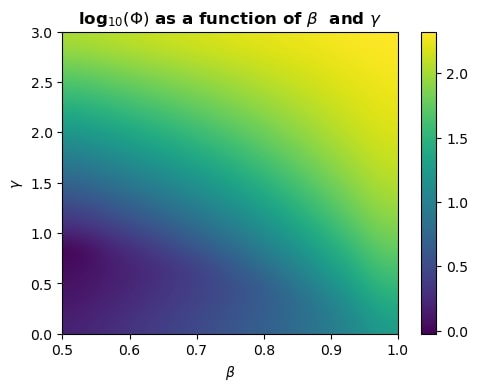}
 \caption{Log of system depletion rate as a function of compression rate $\beta$ and energy capacity density exponent $\gamma$. }
 \label{depletion-rate-beta-gamma}
\end{center}
\end{figure}

Figure~\ref{depletion-rate-beta-gamma} shows how the compression factor $\beta$ and energy capacity density exponent $\gamma$ influence the depletion rate $\Phi$.
It is clear that reducing $\beta$ produces large reductions in the depletion rate, especially when $\gamma$ is small. In the the graph, the smallest value of $\Phi$ attained (corresponding to  $\beta=0.5, \gamma=0$) is over 100 times smaller than the largest value (corresponding to $\beta=1, \gamma = 3$). 
In the default case where $\gamma =1$, then reducing $\beta$ from 1 to 0.5 reduces $\Phi$ by a factor of about 13.

\begin{figure}[!h]
\begin{center}
\includegraphics[width=8cm]{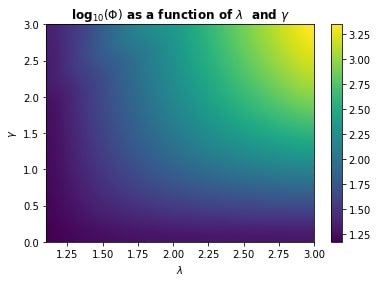}
 \caption{Log of system depletion rate as a function of transmission power exponent $\lambda$ and $\gamma$. }
 \label{depletion-rate-lambda-gamma}
\end{center}
\end{figure}

Figure \ref{depletion-rate-lambda-gamma}  the joint dependence of $\Phi$ on the energy density exponent $\gamma$ and the transmission exponent $\lambda$.
The depletion rate increases with increasing $\lambda$, especially when $\gamma$ is large: for example, when $\gamma = 3$, then the depletion rate for $\lambda = 3$ is about 100 times larger than for $\lambda = 1.1$.  This may be attributed to the fact that when $\lambda$ is increased, long-distance transmission becomes increasingly expensive: and when $\gamma$ is increased, an increasing proportion of the transmission is direct rather than stepwise as shown in \ref{direct-stepwise-total-power-flow} (c). 

\subsection{Depletion rate as a function of coverage area and number of rings}

\begin{figure}[!h]
\begin{center}
\includegraphics[width=8cm]{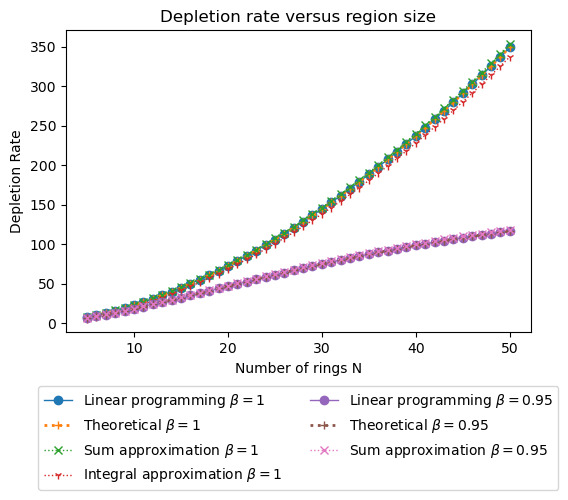}
 \end{center}
 \caption{Depletion rates computed from the linear programming solution, the exact theoretical expressions, the summation approximation and the integral approximation, as a function of number of rings $N$ for $\alpha =\gamma = 1$,  $\lambda = 2$ and $d=1$, and two different compression rates ($\beta = 1$ and $\beta = 0.95$). }
 \label{depletion-rate}
\end{figure}

Figure \ref{depletion-rate} presents the depletion rate as a function of coverage area, where the inter-node distance is fixed at 1, in other words the ring radii increase with increment 1. Two different values of the compression rate $\beta$ are shown, while other parameters take their baseline values. The depletion rate computed using the linear programming formulation in\eqref{eq_propPower}  is compared  with the theoretical equation  \eqref{eq_tPhi} and the two approximations \eqref{eq_PhiOrig} and \eqref{eq_PhiApprox}. The figure shows that depletion rate increases with the number of rings, in good agreement with both approximations. According to \eqref{eq_PhiApprox}, when $\beta \approx 1$ the rate of increase is nearly proportional to $N^{\lambda}$; and when $\beta<1$, the depletion rate tends towards a constant value for large $N$.

\begin{figure}[!h]
\begin{center}
\includegraphics[width=8cm]{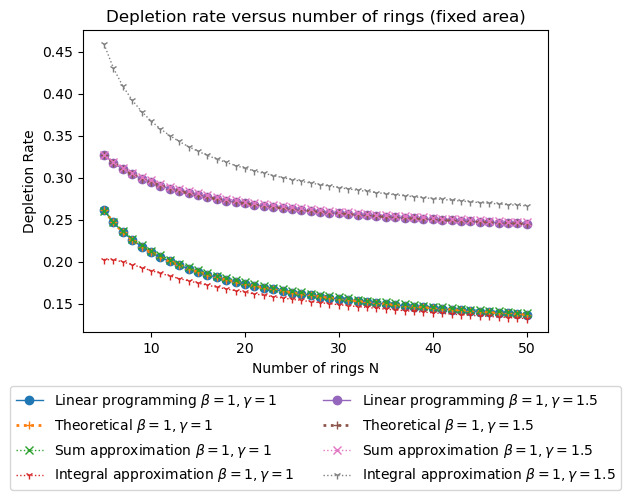}
 \end{center}
 \caption{Depletion rates computed from the linear programming solution, the exact theoretical expressions, the summation approximation and the integral approximation, for different number of rings  $N$ at constant area for $\alpha = \beta =1$, $\lambda = 2$, $d=1/N$, and two different energy density exponents ($\gamma=1$ and $\gamma = 1.5$). }
 \label{depletion-rate-const-area}
\end{figure}

Figure \ref{depletion-rate-const-area} the effect of increasing the number of nodes $N$ (corresponding to the number of rings) while  maintaining a constant coverage area by choosing $d = 1/(N+0.5)$. 
Once again the depletion rate computed using  linear programming  is compared  with the theoretical equation  \eqref{eq_tPhi} and the two approximations \eqref{eq_PhiOrig} and \eqref{eq_PhiApprox}.
Two different values of $\gamma$ are used, while other parameters take baseline values. The figure shows that increasing the number of rings  gradually reduces depletion rate, in very close agreement with \eqref{eq_PhiOrig} and fairly close to \eqref{eq_PhiApprox}, especially for larger values of $N$.  According to the approximation in  
\eqref{eq_PhiApprox}, this decrease goes approximately as $\log N$ when $\gamma=1$, and tends towards a constant value when $\gamma=1.5$. 

\section{Conclusion}
\label{sec6}

In this paper, we 
present a thorough analysis of a particular model of data aggregation for dense WSN's. The model employs a ring-sector division of the coverage area, so that the optimization problem can be reduced to a 1-dimensional model consisting of equally-spaced nodes on a line. This model is optimized using linear programming, and under a wide variety of scenarios the optimal transmission pattern involves only direct and stepwise transmission between nodes (which represent area patches in the ring-sector geometry). When only these two types of transmission are involved, an  analytical solution can be found, which can  approximated by simpler expressions. 

Based of the foregoing analysis and simulations, three strategies for reducing depletion rate (thus increasing system lifetime) may be identified.
These strategies include (1) non-uniform distribution of sensors (reflected by non-constant information density and/or energy capacity density); (2) Using repeated compression to reduce the total information flow; and (3) changing the system geometry by increasing the number of rings, while maintaining the same coverage area.  It was found that strategy (1) has very little effect on system lifetime. In particular, a nonuniform information density had almost no effect on system lifetime, while a power capacity density that is larger near the sink can improve system lifetime by a factor of about 3. Strategy (2) was much more effective, and can lead to system lifetime improvements of more than an order of magnitude. However, this improvement is gained at the cost of information loss: repeated compression implies that a considerable amount of information is discarded during the transmission process, and only a fraction (presumably the most important information) reach the sink. Finally, strategy (3) also was effective, since it was found that increasing the number of  rings reduced the depletion rate. However, increasing the number of rings leads to smaller and smaller patches, so this would need to be balanced against the aggregation capabilities within the patches. 

The limitations of the model should be acknowledged, because in many respects the model is oversimplified. For example, it does not take into account the power consumption required to gather the information within each patch at a single node within the patch. Since this gathering only involves short-range transmission, this power consumption is a secondary effect, but nonetheless appreciable.  Moreover, since the size of the patches increases with ring number, it is increasingly costly (and inefficient) to gather the information from each patch to a single node.  Nonetheless, the model provides insights  that may be useful when examining systems with more sophisticated configurations. 

For future work,  more sophisticated configurations may be considered.  One possibility is to divide the coverage area into equal hexagonal cells, which can be further divided among independent sectors for smart data aggregation. Preliminary investigations show that these cells arrange themselves into patch-like substructures, and that transmission characteristics between patches resemble the characteristics shown in the simple model described in this paper.



\bibliographystyle{elsarticle-num} 
\bibliography{biblio}

\end{document}